\documentclass[prb,twocolumn,floatfix,superscriptaddress,showpacs]{revtex4}
\usepackage{graphicx,color}
\usepackage{amsmath,amssymb,bm,bbm}

\newcommand{\la}{\Lambda}
\newcommand{\tn}{\textnormal}
\newcommand{\cprb}[3]{Phys.~Rev.~B {\bf #1}, #2 (#3)}
\newcommand{\cprl}[3]{Phys.~Rev.~Lett.~{\bf #1}, #2 (#3)}
\newcommand{\cnjp}[3]{New J.~Phys.~{\bf #1}, #2 (#3)}
\newcommand{\cjp}[3]{J.~Phys.: Condensed Matter {\bf #1}, #2 (#3)}

\newcommand{\cbook}[2]{\textit{#1} (#2)}

\definecolor{darkred}{rgb}{0.90,0,0}
\definecolor{darkgreen}{rgb}{0,0.60,.2}
\definecolor{darkblue}{rgb}{0,0,1}
\definecolor{grey}{cmyk}{0,0,0,0.25}
\definecolor{orange}{cmyk}{0,0.6,0.8,0}

\begin{document}
\title{\boldmath Functional renormalization group study\\ of the interacting resonant level model in and out of equilibrium}

\author{C.\ Karrasch}
\affiliation{Institut f\"ur Theoretische Physik A and JARA -- Fundamentals of Future Information Technology, RWTH Aachen University, 52056 Aachen, Germany}
\author{M.\ Pletyukhov}
\affiliation{Institut f\"ur Theoretische Physik A and JARA -- Fundamentals of Future Information Technology, RWTH Aachen University, 52056 Aachen, Germany}
\author{L.\ Borda}
\affiliation{Physikalisches Institut and Bethe Center for Theoretical Physics, Universit\"at Bonn, Nussallee 12, 53115 Bonn, Germany}
\affiliation{Department of Theoretical Physics and Research Group of the Hungarian Academy of Sciences, Budapest University of Technology and Economics, H1111 Budapest, Hungary}
\author{V.\ Meden}
\affiliation{Institut f\"ur Theoretische Physik A and JARA -- Fundamentals of Future Information Technology, RWTH Aachen University, 52056 Aachen, Germany}

\begin{abstract}
We investigate equilibrium and steady-state non-equilibrium transport properties of a spinless resonant level locally coupled to two conduction bands of width $\sim\Gamma$ via a Coulomb interaction $U$ and a hybridization $t'$. In order to study the effects of finite bias voltages beyond linear response, a generalization of the functional renormalization group to Keldysh frequency space is employed. Being mostly unexplored in the context of quantum impurity systems out of equilibrium, we benchmark this method against recently-published time-dependent density matrix renormalization group data. We thoroughly investigate the scaling limit $\Gamma\to\infty$ characterized by the appearance of power laws. Most importantly, at the particle-hole symmetric point the steady-state current decays like $J\sim V^{-\alpha_J}$ as a function of the bias voltage $V\gg t'$, with an exponent $\alpha_J(U)$ that we calculate to leading order in the Coulomb interaction strength. In contrast, we do not observe a pure power-law (but more complex) current-voltage-relation if the energy $\epsilon$ of the resonant level is pinned close to either one of the chemical potentials $\pm V/2$.
\end{abstract}

\maketitle

\section{Introduction}
\label{sec:intro}

Experiments on nanostructures represent a highly-active field of research. Whereas transport properties can be measured straightforward beyond linear response, a theoretical approach to quantum impurities out of equilibrium is challenging in presence of Coulomb interactions which are ubiquitous in low-dimensional systems. Over the last years, a great variety of both numerical as well as analytical methods was developed to study correlation effects on the non-equilibrium dynamics of or steady-state current through quantum dots. Ranking among those are exact Bethe ansatz solutions,\cite{bethe,andrei} perturbative renormalization group schemes,\cite{rtrg1,rtrg2,prg1,prg2} quantum Monte Carlo,\cite{qmc1,qmc2} a real-time path integral approach,\cite{egger} Hamiltonian flow equations,\cite{kehrein} as well as the time-dependent numerical (NRG)\cite{nrg1,nrg2} and density matrix renormalization group (DMRG)\cite{dmrg1,dmrg2,dmrg3,dmrg4} frameworks. While all these methods had long-standing success in computing linear-response properties of quantum impurity systems, the non-equilibrium situation is still a newly-emerging and thus rather unexplored field.

The functional renormalization group (FRG) provides an a priori exact re-formulation of a correlated many-particle problem in terms of coupled flow equations for irreducible vertex functions of arbitrary order.\cite{salmhofer} In the context of quantum dots in equilibrium, even a very simple way to truncate this infinite hierarchy, which can be viewed as a kind of RG enhanced Hartree-Fock approximation, allows for accurately describing the effects of small to intermediate (and sometimes even large) Coulomb interactions $U$ very flexibly and with minor numerical effort. Most importantly, the zero-temperature linear conductance has been computed in good agreement with NRG reference data for a variety of quantum dot geometries.\cite{dotsystems,phaselapses} Employing a more elaborate truncation scheme (where one accounts for the frequency dependence of the two-particle vertex) allows for calculating finite-energy properties like the density of states at least for intermediate values of $U$ (for the single impurity Anderson model).\cite{frequenzen,severinsiam}

In contrast to the regime of linear transport where the strength and limitations of the functional RG in the context of quantum impurity systems have been extensively investigated, there are only few works on the case of non-equilibrium. Even though it was proven possible in the steady-state limit to derive an infinite hierarchy of flow equations in Keldysh frequency space which are structurally identical to those on the Matsubara axis,\cite{severindiplom,gezzi,frgprl} little is known on how different approximation (i.e., truncation) schemes succeed or fail to describe out-of-equilibrium physics of correlated quantum dots. The first aim of this paper is to partly fill this gap by benchmarking functional RG calculations for a very simple impurity model (namely the interacting resonant level model) against recently-published linear-response and time-dependent DMRG data.\cite{dmrgeq,dmrgnoneq}

Quite generally, the interacting resonant level model (IRLM) describes a single localized level (with an energy $\epsilon$) coupled to a bath of delocalized states (featuring a bandwidth $\sim\Gamma$) both by a local Coulomb repulsion $U$ and a hopping matrix element $t'$. It was initially introduced four decades ago to study the equilibrium physics of mixed-valence compounds, and observables were computed by mapping to the anisotropic Kondo model (and using results available for the latter) or by perturbative RG calculations.\cite{irlm1,irlm2,irlm3,irlm4} The two-channel version of the IRLM, which has gained considerable interest within the past few years,\cite{andrei,andreierratum,andreicomment,doyon,laszlo,avi,dmrgnoneq} represents a very simple impurity model to describe charge fluctuations and investigate non-equilibrium transport (driven by a bias voltage $V=(\mu_{L}+\mu_{R})/2$ between the two baths) through a quantum dot. Most notably, accurate time-dependent DMRG data was recently published by Boulat et al.\ (for fairly large values of $t'/\Gamma$) and provides the aforementioned benchmark for the functional RG.\cite{dmrgnoneq} The opposite (so-called scaling) limit of the bandwidth $\Gamma$ being much larger than all other energy scales was characterized by the appearance of universal power laws (by approximate approaches each having its advantages and shortcomings).\cite{dmrgnoneq,doyon,laszlo} In particular, the current through the system was found to decay like $J\sim V^{-\alpha_J(U,\epsilon)}$ for $\Gamma\gg V\gg t'$, both for the impurity energy being small ($\epsilon\ll V$)\cite{doyon,laszlo} or close to one of the chemical potentials ($\epsilon\approx\mu_{L,R}$, $\alpha_J(U,\epsilon\approx\mu_{L,R})=\alpha_J(U,\epsilon=0)/2$).\cite{doyon} Having explored its own strength and drawbacks in comparison with the DMRG reference as well as with new equilibrium NRG data, the functional renormalization group allows for systematically studying the scaling limit of the microscopic IRLM for small to intermediate values of the Coulomb interaction $U$. It is the second aim of this paper to provide a consistent picture of the zero-temperature physics in this parameter regime from the FRG point of view, particularly in relation with prior results.

This paper is organized as follows. We introduce the interacting resonant level model as well as the functional renormalization group approach in non-equilibrium in Secs.~\ref{sec:model} and \ref{sec:method}, respectively. Sec.~\ref{sec:dmrg} is devoted to the comparison of FRG results with time-dependent DMRG data. We systematically investigate the scaling limit in Sec.~\ref{sec:scalinglimit} and conclude with a brief summary. A more elaborate FRG approximation scheme is briefly discussed in the Appendix.

\begin{figure}[t]
\includegraphics[width=0.8\linewidth,clip]{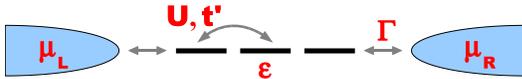}
\caption{(Color online) Schematic presentation of the two-channel interacting resonant level model studied in this paper.}
\label{fig:geometry}
\end{figure}

\section{The Model}
\label{sec:model}

The interacting resonant level model is depicted schematically in Fig.~\ref{fig:geometry}. It describes a single spinless level of energy $\epsilon$ as well as two (left and right) bath of delocalized states:
\begin{equation}\label{eq:himp}
H_\tn{imp} = (\epsilon-U/2) d_2^\dagger d_2^{\phantom{\dagger}}~,~~~H_\tn{bath}^{s=L,R} = \sum_k \epsilon_k^{\phantom{\dagger}} c_{sk}^\dagger c_{sk}^{\phantom{\dagger}}~.
\end{equation}
We model a local Coulomb interaction $U$ and hopping $t'$ between both parts by adding two distinguished neighboring sites,
\begin{equation}\label{eq:hu}\begin{split}
H_\tn{U} =~ U & \left(d_2^\dagger d_2^{\phantom{\dagger}} d_1^\dagger d_1^{\phantom{\dagger}} + d_2^\dagger d_2^{\phantom{\dagger}} d_3^\dagger d_3^{\phantom{\dagger}} \right) \\
 -t' & \left( d_2^\dagger d_1^{\phantom{\dagger}} + d_2^\dagger d_3^{\phantom{\dagger}} + \tn{H.c.} \right)
- U/2\left(d_1^\dagger d_1^{\phantom{\dagger}} + d_3^\dagger d_3^{\phantom{\dagger}}\right)~,
\end{split}\end{equation}
which are coupled to the bath of size $N$ via
\begin{equation}
H_\tn{coup} = -  \frac{t}{\sqrt{N}} \sum_k \left(d_1^\dagger c_{Lk}^{\phantom{\dagger}} + d_3^\dagger c_{Rk}^{\phantom{\dagger}} + \tn{H.c.}\right) ~.
\end{equation}
The characteristic energy scale (i.e., the bandwidth) of the latter is determined by the hybridization
\begin{equation}
\Gamma=\pi\rho_\tn{bath}(\omega=0)t^2~, 
\end{equation}
with $\rho_\tn{bath}(\omega)$ being the local density of states. In order to explicitly compare with DMRG results, we model the baths as semi-infinite tight-binding chains with a nearest-neighbor hopping amplitude $t$ and correspondingly
\begin{equation}
\rho_\tn{bath}(\omega) = \frac{1}{2\pi t^2}\sqrt{4t^2-\omega^2}\ \Theta(2t-|\omega|)~,~~\Gamma=t~.
\end{equation}
The associated retarted Green function at the end of the isolated chain is given by
\begin{equation}
g_\tn{bath}^\tn{ret}(\omega) = \frac{1}{2t^2}
\begin{cases}
 \omega - \tn{sgn}(\omega)\sqrt{\omega^2-4t^2}  & |\omega|>2t \\
\omega - i\sqrt{4t^2-\omega^2} & |\omega|<2t~. \\
\end{cases}
\end{equation}
As an alternative, one frequently employs completely structureless (wide-band) leads featuring a constant local density of states and
\begin{equation}
g_\tn{bath}^\tn{ret}(\omega) = -i\pi\rho_\tn{bath}~.
\end{equation}
In our case, such realization is used to investigate the scaling limit $\Gamma\to\infty$ where details of the dispersion $\epsilon_k$ do not play any role (which one can can show explicitly within the FRG framework; see Sec.~\ref{sec:scalinglimit}).

The equilibrium statistics of the interacting resonant level model is determined by the usual grand canonical density operator $\hat\rho=\exp(-\beta H)$ featuring an inverse temperature $\beta$ and equal chemical potentials $\mu_L=\mu_R=0$. The non-equilibrium situation is modeled by an initially separated system ($H_\tn{coup}=0$) in a thermal bulk state
\begin{equation}
\hat\rho = e^{-\beta H_\tn{bath}^L +\beta\mu_L N_L}
\otimes \hat\rho_\tn{imp} \otimes e^{-\beta H_\tn{bath}^R +\beta\mu_R N_R}~,
\end{equation}
where $\hat\rho_\tn{imp}$ denotes the density matrix of the isolated interacting three-site region, which we choose to be that of a vacuum configuration. At some time $t_0$, the coupling is switched on, and the time evolution for $t>t_0$ is governed by the full Hamiltonian
\begin{equation}\label{eq:h}
H = H_\tn{imp} + H_\tn{U} + H_\tn{coup} + H_\tn{bath}^L + H_\tn{bath}^R~.
\end{equation}
In presence of a finite bias voltage $V=2\mu_L=-2\mu_R$, one does in general expect the system to relax to a non-thermal steady state independent of $\hat\rho_\tn{imp}$ at $t\to\infty$, and this scenario is supported by time-dependent DMRG calculations for the problem at hand.\cite{dmrgnoneq} In this paper, we will focus exclusively on studying the steady state of the IRLM in the zero-temperature limit.

\section{The Method}
\label{sec:method}

\subsection{General idea of the functional RG}
\label{sec:method.general}

The functional renormalization group implements Wilson's general RG idea in terms of an infinite hierarchy of differential flow equations for single-particle irreducible vertex functions (such as the self-energy), which altogether represents an exact reformulation of the underlying many-particle problem.\cite{salmhofer} The hierarchy is set up by introducing an infrared cutoff $\Lambda$ into the non-interacting Green function $G_0$,
\begin{equation}\label{eq:cutoff}
G_0(1';1) \to G_0^\la(1';1)\ ,~~G_0^{\la=\infty}=0\ ,~~G_0^{\la=0}=G_0 \ ,
\end{equation}
where the arguments are a shorthand for single-particle quantum numbers as well as either Matsubara frequencies (in equilibrium) or real frequencies and Keldysh indices (in non-equilibrium). Under the assumption of the existence of a steady state for the latter case (which allows for associating energies with time differences), such an infrared (energy) cutoff can be devised straightfoward, but this will be postponed to the next Section since its actual form is irrelevant for the time being. By virtue of the replacement (\ref{eq:cutoff}), every vertex function acquires a $\la$-dependence, and both in and out of equilibrium one can derive the structurally same set of functional RG flow equations by straight-forwardly differentiating with respect to the cutoff parameter $\la$. This can be technically achieved, e.g., by considering generating functionals, and the flow of the self-energy is given by\cite{dotsystems,severindiplom,gezzi}
\begin{equation}\label{eq:flowse}
\partial_\la\Sigma^\la(1';1) = - \sum_{22'} S^\la(2;2')\gamma_2^\la(1',2';1,2)~,
\end{equation}
with the so-called single-scale propagator
\begin{equation}\label{eq:singlescale}
S^\la = - G^\la\partial_\la\left[G_0^\la\right]^{-1} G^\la = -\partial_\la G^\la~,
\end{equation}
and $G^\la$ being the full Green function at scale $\la$. Both the two-particle vertex $\gamma_2^\la$ and $\Sigma^\la$ itself enter the right-hand side of the differential equation (\ref{eq:flowse}). Similarly, the flow of the $n$-th order function $\gamma_n^\la$ is in general determined by all vertices $\gamma_{i\leq n}^\la$. Integrating this infinite set of coupled differential equations from $\la=\infty$ (where all energy scales are suppressed and the many-particle problem becomes trivial) down to $\la=0$ (where one recovers the full energy spectrum) yields an in principle exact expression for the self-energy of the system under consideration. In practice, however, one needs to devise a truncation scheme. In the main part of this paper, we focus solely on the flow equation (\ref{eq:flowse}) with the two-particle vertex set to its initial value ($\gamma_2^{\la=\infty}\sim U$), rendering $\Sigma$ a frequency-independent (effective) quantity. This Hartree-Fock-like approximation is correct at least to leading order in $U$ and thus a priori justified in the limit of small Coulomb interactions. In equilibrium, it was proven to give reliable results up to intermediate values of $U$ (and even to capture aspects of Kondo physics) for the linear conductance of a variety of quantum dot geometries.\cite{dotsystems,phaselapses} A more elaborate truncation scheme\cite{frequenzen,severinsiam} which accounts for the flow of the frequency-dependent two-particle vertex will be discussed briefly in the Appendix.

\subsection{Green functions \& Dyson equation}
\label{sec:method.gf}

\subsubsection{Green functions in and out of equilibrium}

Linear-response transport properties of the interacting resonant level model can be computed conveniently using Matsubara Green functions
\begin{equation}\begin{split}
G_{ij}^\tn{eq}(i\omega) & = - \int_0^\beta e^{i\omega\tau}\left\langle T_\tau a_i^{\phantom{\dagger}}(\tau) a_j^\dagger(0)\right\rangle d\tau \\
& = \left[G_0^\tn{eq}(i\omega)^{-1} - \Sigma^{\tn{eq}}(i\omega)\right]^{-1}_{ij}~,
\end{split}\end{equation}
where $a_i(\tau)$ is a fermionic annihilation operator in the Heisenberg picture, $T_\tau$ denotes ordering with respect to the imaginary time $\tau$, and the self-energy $\Sigma^\tn{eq}$ is associated with the Coulomb interaction. In order to employ standard diagrammatic techniques in non-equilibrium (such as the very notion of vertex functions), we define long-time Green functions as
\begin{equation}\begin{split}
\hat G_{ij}(\omega) & = -i\int e^{i\omega t}\lim_{t_0\to-\infty}\left\langle T_c a_i^{\phantom{\dagger}}(t) a_j^\dagger(0)\right\rangle_{\hat\rho} dt \\
& =  \left[\hat G_0(\omega)^{-1} - \hat\Sigma(\omega)\right]^{-1}_{ij} = \begin{pmatrix}
G^{--}_{ij}(\omega) & G^{-+}_{ij}(\omega) \\ G^{+-}_{ij}(\omega) & G^{++}_{ij}(\omega)
\end{pmatrix},
\end{split}\end{equation}
with $T_c$ being the order operator on the Keldysh contour whose branches are characterized by indices $\alpha=\pm$. It is often more convenient to directly exploit causality (particularly in approximate schemes which naturally conserve this symmetry; see Sec.~\ref{sec:method.frg}) and work in the retarded, advanced, and Kelydsh basis:
\begin{equation}\begin{split}
G^\tn{ret} &=  G^{--} - G^{-+} = \left(G^\tn{adv}\right)^\dagger,~
G^\tn{K}  =  G^{-+} + G^{+-}~, \\
\Sigma^\tn{ret} &=  \Sigma^{--} + \Sigma^{-+} = \left(\Sigma^\tn{adv}\right)^\dagger,~
\Sigma^\tn{K}  =  \Sigma^{--} + \Sigma^{++}~.
\end{split}\end{equation}
In the next Section, we will explicitly derive the (non-interacting) Green functions of the IRLM in and out of equilibrium.

\subsubsection{Dyson equation}

For the problem at hand, the flow of the self-energy is determined by the Green functions of the interacting three-site system only. Using standard projection techniques,\cite{taylor} the latter can be expressed in terms of a finite matrix Dyson equation
\begin{equation}\begin{split}
G^\tn{ret}(\omega)^{-1} & \stackrel{\phantom{\tn{equil.}}}{=}~ G^\tn{ret}_0(\omega)^{-1} -\Sigma^\tn{ret}(\omega) \\
& \stackrel{\phantom{\tn{equil.}}}{=}~ g^\tn{ret}_\tn{imp}(\omega)^{-1} - \Sigma_\tn{bath}^\tn{ret}(\omega)  - \Sigma^\tn{ret}(\omega)\\
& \stackrel{\tn{equil.}}{=}~ G^\tn{eq}(i\omega\to\omega+i0)^{-1} \\[1ex]
G^\tn{K}(\omega) & \stackrel{\phantom{\tn{equil.}}}{=}~  G^\tn{ret}(\omega)\left[\Sigma^\tn{K}(\omega) + \Sigma^\tn{K}_\tn{bath}(\omega)\right]G^\tn{adv}(\omega)~,
\end{split}\end{equation}
where the retarded Green function of the isolated impurity region is given by
\begin{equation}
g^\tn{ret}_\tn{imp}(\omega)^{-1} =
\begin{pmatrix}
\omega+i0 & -t' & 0 \\ -t' & \omega-\epsilon + i0 & -t' \\ 0 & -t' & \omega+i0
\end{pmatrix}~,
\end{equation}
and the self-energy associated with the bath reads
\begin{equation}\begin{split}
\Sigma_\tn{bath}^\tn{ret}(\omega) & = t^2 g_\tn{bath}^\tn{ret} (\omega)
\begin{pmatrix}
 1 & & \\ & 0 & \\ & & 1
\end{pmatrix}~, \\
\Sigma^\tn{K}_\tn{bath}(\omega) & = -2i\pi \rho_\tn{bath}t^2
\begin{pmatrix}
1-2f_L & & \\ & 0 & \\ & & 1-2f_R
\end{pmatrix}.
\end{split}\end{equation}
The latter are initially in a thermal state (for which the dissipation-fluctuation theorem holds) and described by the Fermi functions
\begin{equation}
f_{L,R}(\omega) = \frac{1}{e^{(\beta-\mu_{L,R})\omega}+1}~.
\end{equation}
In the next Section, we will derive the functional renormalization group flow equations in order to compute an approximation for the self-energies $\Sigma^\tn{ret}$ and $\Sigma^\tn{K}$ (out of equilibrium) or $\Sigma^\tn{eq}$ (for linear-response) which in each case incorporate the effects of the Coulomb repulsion. Afterwards, one can calculate the current flowing through the interacting region as well as the zero-temperature equilibrium conductance using the formulas (in units of $e^2/h=1$)\cite{wingreen}
\begin{equation}\label{eq:jg}\begin{split}
J_{s} & = 2\pi i t^2\int \rho_\tn{bath}\left[f_s\left(G^{+-}_{ii}-G^{-+}_{ii} \right)+G^{-+}_{ii}\right] d\omega~,\\
G & = 4\Gamma^2\left|G^\tn{eq}_{13}(i\omega=0)\right|^2~,
\end{split}\end{equation}
with the Green function's single-particle index being $i=1$ or $i=3$ for the current at the left and right interface ($s=L,R$), respectively. Within all FRG approximation schemes employed to study the problem at hand, current conservation $J_L=-J_R$ holds (whereas other symmetries may be violated; see Sec.~\ref{sec:method.frg}).

\subsection{Flow equations}
\label{sec:method.frg}

In this paper, we will almost exclusively focus on considering the flow of the self-energy only (i.e., truncating the infinite RG hierarchy to leading order). This is achieved by setting the two-particle vertex $\gamma_2^\la$ to its initial value in Eq.~(\ref{eq:flowse}), which in equilibrium is nothing but the bare frequency-independent Coulomb interaction
\begin{equation}\label{eq:initialvertexeq}\begin{split}
& \gamma_2^\la(1',2';1,2) = \pm U \beta^{-1} \delta(i\omega_{1'}+i\omega_{2'}-i\omega_1-i\omega_2)
\end{split}\end{equation}
for all permutations of the nearest-neighbor single-particle indices. An additional factor of imaginary $i$ appears in a real frequency representation, and the two-particle vertex is only non-vanishing if all Kelydsh indices $\alpha_{1'}=\alpha_{2'}=\alpha_{1}=\alpha_{2}=\alpha$ are equal:\cite{severindiplom}
\begin{equation}\label{eq:initialvertex}
\gamma_2^\la(1',2';1,2) = \pm \alpha i U \delta(\omega_{1'}+\omega_{2'}-\omega_1-\omega_2)~.
\end{equation}
As in Eq.~(\ref{eq:initialvertexeq}), different signs refer to symmetric or antisymmetric ways of ordering nearest-neighbor quantum numbers. As mentioned above, the self-energy obtained from this truncation scheme is frequency-independent (i.e., Hartree-Fock-like) and at least correct to first order in $U$. It contains, however, certain classes of higher-order contributions due to the underlying RG procedure and was shown to be reliable for small to intermediate Coulomb interactions and even to capture non-perturbative aspects of strong electronic correlations in the context of equilibrium quantum dot systems.\cite{dotsystems,phaselapses} A more elaborate approximation scheme which accounts for the frequency-dependence of the two-particle vertex\cite{frequenzen,severinsiam} will be discussed briefly in the Appendix (for the IRLM in linear response).

\subsubsection{Equilibrium}
\label{sec:method.equil}

The last step in explicitly setting up the functional RG flow equations is to specify the form of the infrared cutoff. In equilibrium, low-energy degrees of freedom are most commonly suppressed by a sharp multiplicative $\Theta$-function in Matsubara frequency space:
\begin{equation}\label{eq:cutoffeq}
G_0^\tn{eq}(i\omega) \to G_0^{\tn{eq},\la}(i\omega) = \Theta(|\omega|-\la) G_0^\tn{eq}(i\omega)~,
\end{equation}
and we will use such implementation throughout this paper in order to compute linear-response properties of the IRLM. The single-scale propagator $S^{\tn{eq},\la}$, which solely determines the right-hand side of Eq.~(\ref{eq:flowse}) for a two-particle vertex set to its initial value, contains an at first sight ambiguous product of $\Theta$- and $\delta$-functions. Evaluating the latter by means of Morris lemma\cite{morris} yields
\begin{equation}
S^{\tn{eq},\la} = \delta(|\omega|-\la)\left[\left(G_0^\tn{eq}\right)^{-1} \hspace*{-0.3ex}- \Sigma^{\tn{eq},\la}\right]^{-1}
\hspace*{-0.3ex}= \delta(|\omega|-\la) \tilde G^{\tn{eq}, \la},
\end{equation}
and the zero-temperature flow equations for the different independent self-energy components $\Sigma_{12}^{\tn{eq},\la}+t'=t'_\la$, $\Sigma_{22}^{\tn{eq},\la}+\epsilon=\epsilon_\la$, and $\Sigma_{11}^{\tn{eq},\la}=\epsilon'_\la$  (introducing a notation where the interpretation as effective system parameters becomes evident) read
\begin{flalign}\label{eq:floweqts}
\partial_\la t'_\la & = \phantom{-}\frac{U}{\pi}\, \tn{Re}\left[\tilde G^{\tn{eq}, \la}_{12}(i\la)\right] \hspace*{-0.5cm} & 
t'_{\la\to\infty} & =t' \\
\partial_\la \epsilon_\la & = -\frac{U}{\pi}\, \tn{Re}\left[\tilde G^{\tn{eq}, \la}_{11}(i\la)+\tilde G^{\tn{eq}, \la}_{33}(i\la)\right] \hspace*{-0.5cm}
& \epsilon_{\la\to\infty}& =\epsilon \label{eq:floweqe} \\
\partial_\la \epsilon'_\la & = -\frac{U}{\pi}\, \tn{Re}\left[\tilde G^{\tn{eq}, \la}_{22}(i\la)\right] \hspace*{-0.5cm}
& \epsilon'_{\la\to\infty}& =0 \label{eq:floweqes}
\end{flalign}
The single-particle energy shift $-U/2$ appearing in the Hamiltonian of Eqs.~(\ref{eq:himp}) and (\ref{eq:hu}) is cancelled by a contribution arising from analytically integrating from $\la=\infty$ down to some arbitrarily large $\la\to\infty$. The ordinary coupled differential equations (\ref{eq:floweqts}) - (\ref{eq:floweqes}) can be solved numerically (and sometimes even exactly; see Sec.~\ref{sec:scalinglimit}) with minor effort, and we will discuss the linear-response physics described by this approximation scheme in Secs.~\ref{sec:dmrg} and \ref{sec:scalinglimit}.

\subsubsection{Non-equilibrium: sharp cutoff scheme}
\label{sec:method.theta}

One straightforward way of introducing a cutoff in non-equilibrium is the replacement analogous to Eq.~(\ref{eq:cutoffeq}):
\begin{equation}\label{eq:thetacutoff}
\hat G_0(\omega) \to \hat G_0^\la(\omega) = \Theta(|\omega|-\la)
\begin{pmatrix}
G_0^{--}(\omega) & G_0^{-+}(\omega) \\ G_0^{+-}(\omega) & G_0^{++}(\omega)
\end{pmatrix}~.
\end{equation}
As before, the single-scale propagator $S^\la$ which enters on the right-hand side of Eq.~(\ref{eq:flowse}) can be calculated by virtue of Morris lemma:\cite{morris}
\begin{equation}
\hat S^\la = \delta(|\omega|-\la)\left[\hat G_0^{-1} - \hat\Sigma^\la\right]^{-1}
= \delta(|\omega|-\la) \hat{\tilde G}^\la~,
\end{equation}
and the flow of the self-energy is given by
\begin{equation}\label{eq:flowsenoneq}\begin{split}
\partial_\la \Sigma^{--, \la}_{22} & = \phantom{-}\frac{iU}{2\pi}\sum_{\omega=\pm\la}\Big[\tilde G^{--, \la}_{11}(\omega) + \tilde G^{--, \la}_{33}(\omega)\Big] \\
\partial_\la \Sigma^{--, \la}_{11} & = \phantom{-}\frac{iU}{2\pi}\sum_{\omega=\pm\la}\phantom{\Big[}\tilde G^{--, \la}_{22}(\omega) = \partial_\la \Sigma^{--, \la}_{33} \\
\partial_\la \Sigma^{--, \la}_{ij} & = -\frac{iU}{2\pi}\sum_{\omega=\pm\la}\phantom{\Big[}\tilde G^{--, \la}_{ij}(\omega)
\, ,~(ij)=(12,21,23,32)\\
\partial_\la\Sigma^{++, \la} & = - \left(\partial_\la\Sigma^{--, \la}\right)^\dagger \\[1ex]
\partial_\la\Sigma^{+-, \la} & = \partial_\la\Sigma^{-+, \la} = 0~,
\end{split}\end{equation}
complemented by the initial condition $\hat\Sigma^{\la\to\infty}=0$. Since non-equilibrium symmetry properties (particularly causality; see below) are not necessarily conserved by this approximation, it is not reasonable (but misleading) to interpret the different frequency-independent self-energy components as effective non-interacting system parameters.

The very same sharp cutoff scheme defined by Eq.~(\ref{eq:thetacutoff}) has been previously applied to the single impurity Anderson model (SIAM).\cite{gezzi} While qualitatively reproducing non-equilibrium features known, e.g., from perturbation theory, its major drawback (besides numerical problems) turned out to be the above-mentioned symmetry violations. Most importantly, the causality condition
\begin{equation}
\Sigma^{++}+\Sigma^{--}+\Sigma^{-+}+\Sigma^{+-} = 0
\end{equation}
only holds to the truncation order (i.e., to leading $U$ in the present context). Since the functional RG is a generally non-conserving approximation,\cite{hartreefock} the consequences of these violations of symmetries specifically associated with non-equilibrium are a priori unclear. Due to the lack both of reliable reference data as well as of an alternative idea to introduce a conserving cutoff procedure, it was not possible to systematically address this question for the Anderson model at that point of time. Only recently,\cite{severinsym} Jakobs et al.~introduced a scheme to suppress low-energy degrees of freedom (presented in the next Section) which does not violate causality in non-equilibrium (but features other shortcomings\cite{reservoirproblem}), and thorough investigations of the SIAM are under way.\cite{severinsiam} In this paper, we study the interacting resonant level model using both FRG cutoff schemes, particularly in comparison with accurate DMRG results.\cite{dmrgeq,dmrgnoneq}

\begin{figure*}[t]
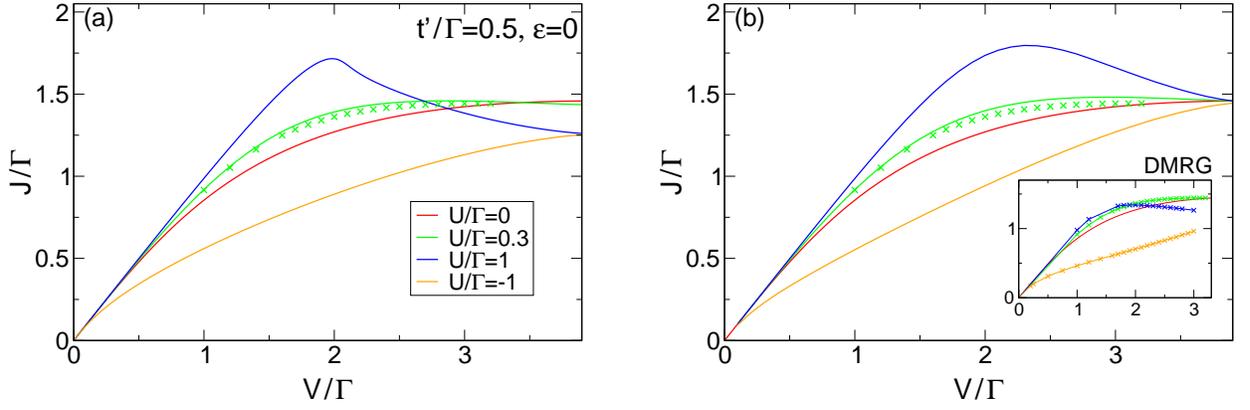

\includegraphics[height=5.3cm,clip]{dmrg1.eps}\hspace*{0.05\linewidth}
\includegraphics[height=5.3cm,clip]{dmrg2.eps}
\caption{(Color online) The steady-state current $J$ as a function of the bias voltage $V$ of the two-channel IRLM for large hoppings $t'=0.5\Gamma$ (in units of the bandwidth $\sim\Gamma$), zero impurity energy $\epsilon$, and various Coulomb interactions $U$. (a) Functional renormalization group results obtained from numerical integration of the self-energy flow equations (\ref{eq:flowsenoneq}) of the sharp cutoff scheme of Sec.~\ref{sec:method.theta}. (b) The same calculated with the reservoir cutoff approach of Sec.~\ref{sec:method.gamma}. Density-matrix renormalization group data of Ref.~\onlinecite{dmrgnoneq} for the same set of parameters is shown by symbols in the main part (for $U=0.3\Gamma$ only) as well as within the inset (where lines are a guide to the eye only).}
\label{fig:dmrgj}
\end{figure*}

\subsubsection{Non-equilibrium: reservoir cutoff scheme}
\label{sec:method.gamma}

The aforementioned alternative way to cut off low energy modes within the functional renormalization group can be introduced on the Hamiltonian level as additional structureless reservoirs of zero chemical potential locally coupled to each site of the interacting region
\begin{equation}
H_\tn{cut} = -\frac{t_\la}{\sqrt{N}}\sum_{i=1}^3\sum_kd_i^\dagger f_{ik}^{\phantom{\dagger}} + \tn{H.c.}~,
\end{equation}
where the hybridization $\la=\pi\rho_\tn{cut} t_\la^2$ is used as the flow parameter. As we will show later on, this cutoff scheme preserves causality even after truncation, rendering it reasonable to directly work with retarded, advanced and Keldysh Green functions. The latter acquire a new self-energy-like term due to the additional reservoirs
\begin{equation}
\Sigma_\tn{cut}^\tn{ret} = -i\la\mathbbm{1}_3~,~~~\Sigma^\tn{K}_\tn{cut} = -2i\la \tanh(\beta\omega/2)\mathbbm{1}_3~,
\end{equation}
and the corresponding single-scale propagators of Eq.~(\ref{eq:singlescale}) read
\begin{equation}\begin{split}
S^{\tn{ret}, \la} & = i G^{\tn{ret}, \la}G^{\tn{ret}, \la} = \left(S^{\tn{adv}, \la}\right)^\dagger \\[1ex]
S^{\tn{K}, \la} & =- \partial_\la \left[G^{\tn{ret}, \la}\left(\Sigma^{\tn{K}, \la}+\Sigma^\tn{K}_\tn{bath}+\Sigma^\tn{K}_\tn{cut} \right)G^{\tn{adv}, \la}\right] \\
& =   \phantom{+} S^{\tn{ret}, \la}\left(\Sigma^{\tn{K}, \la}+\Sigma^\tn{K}_\tn{bath}+\Sigma^\tn{K}_\tn{cut} \right)G^{\tn{adv}, \la}\\
& \phantom{=} + G^{\tn{ret}, \la}\left(\Sigma^{\tn{K}, \la}+\Sigma^\tn{K}_\tn{bath}+\Sigma^\tn{K}_\tn{cut} \right)S^{\tn{adv}, \la}\\
&\phantom{=}+2i\tanh(\beta\omega/2) G^{\tn{ret}, \la}G^{\tn{adv}, \la}~.
\end{split}\end{equation}
The zero-temperature flow of the effective system parameters
\begin{equation}\begin{split}
t'^\la_{12} -t'& = \Sigma^{\tn{ret}, \la}_{12} = \left(\Sigma^{\tn{adv}, \la}_{21}\right)^*
= \left(\Sigma^{\tn{ret}, \la}_{21}\right)^*  \\
t'^\la_{23}-t' & = \Sigma^{\tn{ret}, \la}_{23} = \left(\Sigma^{\tn{adv}, \la}_{32}\right)^*
= \left(\Sigma^{\tn{ret}, \la}_{32}\right)^*  \\
\epsilon_\la-\epsilon & = \Sigma^{\tn{ret}, \la}_{22} = \left(\Sigma^{\tn{adv}, \la}_{22}\right)^*
= \left(\Sigma^{\tn{ret}, \la}_{22}\right)^*  \\
\epsilon'_\la & = \Sigma^{\tn{ret}, \la}_{11} = \left(\Sigma^{\tn{adv}, \la}_{11}\right)^*
= \left(\Sigma^{\tn{ret}, \la}_{11}\right)^* = \Sigma^{\tn{ret}, \la}_{33}
\end{split}\end{equation}
can be derived straightforward by plugging the constant Coulomb interaction vertex given by Eq.~(\ref{eq:initialvertex}) into the general flow equation (\ref{eq:flowse}) and rotating to the Keldysh basis. One obtains
\begin{flalign}
\partial_\la t'^\la_{12} \hspace*{-0.05cm}& = -\frac{iU}{4\pi}\hspace*{-0.05cm}\int\hspace*{-0.11cm}\phantom{\Big[} S^{\tn{K}, \la}_{12}(\omega)\, d\omega
\hspace*{-0.5cm}& t'^{\la\to\infty}_{12}&=t'  \label{eq:flownoneqts1}\\
\partial_\la t'^\la_{23} \hspace*{-0.05cm}& = -\frac{iU}{4\pi}\hspace*{-0.05cm}\int\hspace*{-0.11cm}\phantom{\Big[} S^{\tn{K}, \la}_{23}(\omega)\, d\omega
\hspace*{-0.5cm}& t'^{\la\to\infty}_{23}&=t' \label{eq:flownoneqts2}\\
\partial_\la \epsilon_\la \hspace*{-0.05cm}& = \phantom{-}\frac{iU}{4\pi}\hspace*{-0.05cm}\int\hspace*{-0.11cm} \left[S^{\tn{K}, \la}_{11}(\omega)+S^{\tn{K}, \la}_{33}(\omega)\right]\hspace*{-0.03cm}d\omega \hspace*{-0.5cm}& \epsilon_{\la\to\infty}&=\epsilon 
\label{eq:flownoneqe}\\
\partial_\la \epsilon'_\la \hspace*{-0.05cm}& = \phantom{-}\frac{iU}{4\pi}\hspace*{-0.05cm}\int\hspace*{-0.11cm}\phantom{\Big[} S^{\tn{K}, \la}_{22}(\omega)\, d\omega
\hspace*{-0.5cm} & \epsilon'_{\la\to\infty} &=0. \label{eq:flownoneqes}
\end{flalign}
We note that $\Sigma^{\tn{ret}, \la}=\Sigma^{\tn{adv}, \la}$ only holds within the first order FRG approach used in this paper. Another characteristic of the latter is that the Kelydsh component of the self-energy does not flow:
\begin{equation}\begin{split}
\partial_\la\Sigma^{\tn{K}, \la} & = \pm\frac{iU}{4\pi}\int\left[S^{\tn{ret}, \la}(\omega)+S^{\tn{adv}, \la}(\omega)\right]d\omega \\
& = \begin{cases} i\pi - i\pi=0 & \tn{\phantom{off-}diag.~components}\\ 0 & \tn{off-diag.~components} , \end{cases}
\end{split}\end{equation}
and the very same holds if one formally considers the `anti-causal' self-energy:
\begin{equation}
\partial_\la\left(\Sigma^{++, \la} + \Sigma^{--, \la} + \Sigma^{-+, \la} + \Sigma^{+-, \la}\right) = 0~.
\end{equation}
Thus, causality is not violated by this FRG approximation scheme, providing the a posteriori justification to work in the basis of retarted, advanced, and Keldysh Green functions. The conservation of other symmetry properties (particularly of the current) follows straight-forwardly from the interpretation of the self-energy as effective (non-interacting) system parameters.\cite{reservoirproblem}

\subsubsection{From non-equilibrium to linear-response}

As mentioned above, we compute linear-response properties of the IRLM using the Matsubara functional RG. The non-equilibrium formalism, however, is applicable for arbitrary bias voltages $V$ and can thus be used to (approximately) describe equilibrium physics in the limit $V\to0$. The latter is particularly simple within the reservoir cutoff scheme introduced in Sec.~\ref{sec:method.gamma}. Namely, at $V=0$ the self-energy flow equation can be rewritten as (schematically omitting single-particle quantum numbers)
\begin{equation}\label{eq:eqnoneq}\begin{split}
\partial_\la\Sigma^{\tn{ret}, \la} & = \mp \frac{iU}{4\pi}\int \partial_\la G^{\tn{K}, \la}(\omega) d\omega \\
& = \pm \frac{U}{4\pi}\int\tanh\left(\frac{\beta\omega}{2}\right)\partial_\omega\left[G^{\tn{ret}, \la}+G^{\tn{adv}, \la}\right]d\omega \\
& = \pm\frac{U}{2\pi}\left[G^{\tn{ret}, \la}(0)+G^{\tn{adv}, \la}(0)\right] \\
& = \pm\frac{U}{\pi}\,\tn{Re}\, G^{\tn{eq}, \la} (i\la)~,
\end{split}\end{equation}
where we have used the fluctuation-dissipation theorem at zero temperature and assumed structureless wide-band leads with a frequency-independent self-energy given by $\pm i\la$, the latter allowing for the replacement $\partial_\la\to\pm i\partial_\omega$. The resulting approximate flow equation (\ref{eq:eqnoneq}), however, is nothing but the Matsubara result at the same order of truncation.\cite{gammaeq} In contrast, it is in general impossible to analytically analyze the flow within the sharp cutoff scheme in the linear-response limit similarly to Eq.~(\ref{eq:eqnoneq}), and the same holds for the situation of baths featuring an energy-dependent local density of states. For the problem at hand, we have numerically studied the case of $V=0$ for those cases in order to ensure that the results are always in agreement with the Matsubara formalism.

\begin{figure}[t]
\includegraphics[height=5.3cm,clip]{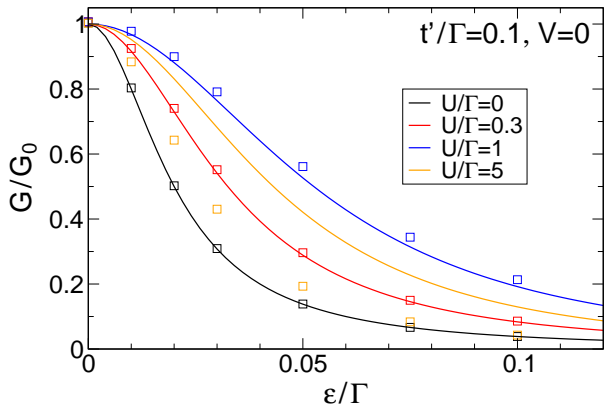}
\caption{(Color online) Linear-response conductance (in units of $G_0=e^2/h$) of the IRLM at $t'=0.1\Gamma$ as a function of the gate voltage $\epsilon$. The Figure shows a comparison between functional RG data obtained from the formalism of Sec.~\ref{sec:method.equil} (lines) and the DMRG results of Ref.~\onlinecite{dmrgeq} (symbols). }
\label{fig:dmrgg}
\end{figure}

\section{Comparison with DMRG}
\label{sec:dmrg}

In this Section, we show results both for the linear-response conductance and the steady-state current of the interacting resonant level model. We focus on the parameter regime of large values of $t'/\Gamma$ (particularly in non-equilibrium) for which reliable linear-response and time-dependent DMRG data was published recently.\cite{dmrgeq,dmrgnoneq} In order to explicitly compare with these results, we model our bath as infinite tight binding chains of bandwidth $4\Gamma$. The so-called scaling limit of $\Gamma\to\infty$ will be discussed extensively in Sec.~\ref{sec:scalinglimit}.

\subsection{Steady-state current}
\label{sec:dmrg.j}

The current $J$ flowing between the two bath of the IRLM in presence of a finite bias voltage $V$ is shown in Figs.~(\ref{fig:dmrgj}a) and (b) for different values of the Coulomb interaction $U$, at the particle-hole symmetric point $\epsilon=0$, and for fixed large hoppings $t'=0.5\Gamma$. It was obtained from numerically integrating the non-equilibrium flow equations (\ref{eq:flowsenoneq}) for the $\Theta$-approach (Fig.~\ref{fig:dmrgj}(a)) and (\ref{eq:flownoneqts1}) - (\ref{eq:flownoneqes}) for the reservoir cutoff scheme (Fig.~\ref{fig:dmrgj}(b)) as well the formula (\ref{eq:jg}), respectively. At $U=0$, $J$ increases linearly for small bias voltages and saturates beyond some scale which is determined by $t'$ (and will be quantified in Sec.~\ref{sec:scalinglimit}). In presence of a finite repulsive Coulomb interaction, it additionally features a regime $V\gtrsim t'^2/\Gamma$ of negative differential conductance (i.e., a current decreasing as the voltage is increased) which was frequently described in previous works (but still lacks a consistent physical explanation).\cite{andrei,dmrgnoneq,doyon,laszlo} Since the hopping $t'$ is fairly large, this decay is not governed by any specific (e.g., power-law-like) form. Most importantly, both FRG schemes show a satisfying agreement with DMRG data both for repulsive and attractive Coulomb interactions (see the symbols in the main part of Figs.~\ref{fig:dmrgj}(a) and (b) as well as the inset to the latter), and the violation of causality prone to the sharp cutoff does not lead to unphysical results. This indicates that even the most simple functional RG truncation scheme captures aspects of the essential non-equilibrium physics of the IRLM, giving confidence to use this approach in order to investigate the current-voltage-relation more thoroughly in the so-called scaling limit $\Gamma\to\infty$ which cannot be accessed straightforward within the time-dependent DMRG framework (see Sec.~\ref{sec:scalinglimit}).

\subsection{Linear response}
\label{sec:dmrg.g}

As a next step, we study the conductance $G$ of the IRLM in the limit of linear response (see Fig.~\ref{fig:dmrgg}) using the equilibrium FRG introduced in Sec.~\ref{sec:method.equil}. In the non-interacting case, its gate voltage dependence is given by a Lorentzian curve whose width is governed by $t'$ (and shows power-law behavior if the latter becomes small compared to $\Gamma$; see Sec.~\ref{sec:scalinglimit}). In presence of small Coulomb interactions, the conductance is enhanced (and the Lorentzian widens) but eventually shrinks as $U$ becomes large. This effect was first observed in equilibrium DMRG calculations,\cite{dmrgeq} and our approximate FRG data agrees quantitatively with those numerically exact results up to up $U/\Gamma\approx1$ and at least qualitatively for even larger interactions.

In order to quantitatively compare with the DMRG data, we employed the IRLM with tight-binding leads featuring an energy-dependent local density of states. In that case, it is a priori impossible to analytically analyze the non-equilibrium FRG flow equations of Sec.~\ref{sec:method.gamma} in the linear-response limit, and the same holds for the sharp cutoff scheme in any case. We have therefore checked numerically that the linear-response conductance obtained from both non-equilibrium frameworks agrees with that of the Matsubara formalism (the latter being shown in Fig.~\ref{fig:dmrgg}).

All in all, the comparison with DMRG data establishes the very simple (Hartree-Fock-like) FRG approach as a satisfying tool to investigate transport properties of the interacting resonant level model at small to intermediate Coulomb interactions in and out of equilibrium.

\begin{figure}[t]
\includegraphics[height=5.1cm,clip]{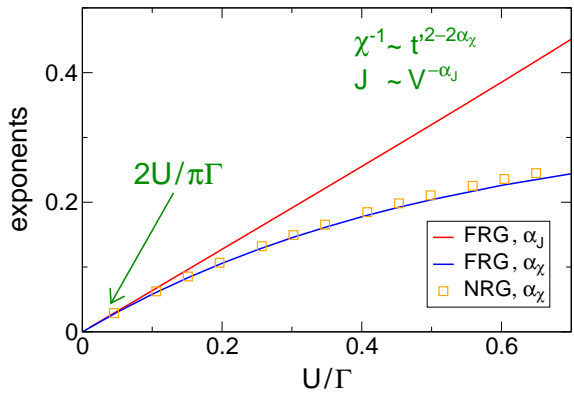}
\caption{(Color online) The exponents $\alpha_J$ and $\alpha_\chi$ governing the scaling-limit power-law behavior of the current (for $t'\ll V$) and of the susceptibility (for $V\ll t'$), respectively. To leading order, both quantities are given by $\alpha=2U/\pi\Gamma$. For the equilibrium exponent $\alpha_\chi$, symbols show numerical renormalization group reference data.  }
\label{fig:exp}
\end{figure}

\section{Scaling limit}
\label{sec:scalinglimit}

In this Section, we investigate the situation where the characteristic energy $\Gamma$ of the bath of delocalized states (i.e., the bandwidth) is much larger than all other energy scales. This so-called scaling limit was addressed in several prior works (with a special focus to non-equilibrium) and is supposed to be governed by universal power laws.\cite{doyon,laszlo} We will particularly discuss the predictions of our FRG approximation scheme in relation with those earlier results.

In the limit of large $\Gamma$, universal features (such as the exponents governing power laws) should only depend on the density of states at the chemical potential but not on the precise energetic structure of the bath. We have numerically checked that this indeed holds at least within our FRG approximation and will thus for simplicity model the leads as completely structureless throughout this Section.

\subsection{Renormalization of $t'$}
\label{sec:scaling.ts}

It is instructive to study the renormalization of the system parameters before discussing physical quantities. In the most simple case of equilibrium and zero impurity energy $\epsilon=0$, the only remaining flow equation (\ref{eq:floweqts}) for of the hopping amplitude $t'^\la$ takes the form
\begin{equation}\label{eq:flowts2}
\partial_\la t'^\la = -\frac{U}{\pi\Gamma} \frac{t'^\la/\Gamma}{(\la/\Gamma)^2+\la/\Gamma+2(t'^\la/\Gamma)^2}~,
\end{equation}
with the initial condition given by $t'^{\la\to\infty}=t'$. Thus, the flow of $t'^\la$ starts below some ultraviolet scale $\la\sim\Gamma$ and is cut by itself in the infrared regime. More precisely, the differential equation (\ref{eq:flowts2}) can be solved analytically in the limit of small $U/\Gamma$ and $t'/\Gamma$ by setting $t'^\la$ in the denominator to its initial value. The result reads
\begin{equation}
\left(\frac{t'^{\la=0}}{t'}\right)^2 \sim \left(\frac{t'}{\Gamma}\right)^{-\frac{4U}{\pi\Gamma}}~,
\end{equation}
and the renormalized hopping scales as a power law with its bare value. Higher-order corrections to the exponent (which in contrast to the first order do not necessarily need to have the right prefactors as our truncated FRG is only correct to leading $U$) can be determined by numerically integrating the full Eq.~(\ref{eq:flowts2}) and fitting a power law to the solution $t'^{\la=0}(t')$. As a side remark, we note that calculating the self-energy by first order perturbation theory in $U$ (which can be most simply done by completely discarding the feedback of $t'^\la$ on the right-hand side of Eq.~(\ref{eq:flowts2})) yields a logarithmic divergence in the bare system parameter,\cite{logdiv}
\begin{equation}
\left(\frac{t'^{\,\tn{pert}}}{t'}\right)^2 = 1 - \frac{4U}{\pi\Gamma}\,\ln\left(\frac{t'}{\Gamma}\right)~,
\end{equation}
indicating the necessity to employ an RG-based framework.

Using similar analytical arguments at finite impurity energies and out of equilibrium, one can show that to leading order in $U$ the flow equations for $\tau_{1,3}^\la=|t'^{\la}_{12,23}|^2/\Gamma$ take the general form\cite{noneqcomment1}
\begin{equation}\label{eq:flowts3}
\partial_\la \tau_{1,3}^\la = -\tau_{1,3}^\la\frac{2U}{\pi\Gamma}\frac{\la+\tau_1^{\la}+\tau_3^{\la}}{(V/2\mp\epsilon)^2+(\la+\tau_1^{\la}+\tau_3^{\la})^2}~,
\end{equation}
for $\la$ smaller than the ultraviolet cutoff $\sim\Gamma$. Thus, the effective hopping amplitudes $\tau_{1,3}=\tau_{1,3}^{\la=0}$ scale as a power law with each of the bare system parameters $V$, $\epsilon$, and $t'$, provided that particular one is much larger (the meaning of which will be quantified in the next Section) than the other two but still much smaller than the bandwidth $\Gamma$:
\begin{equation}\label{eq:tren}
\frac{\tau_{1,3}}{t'^{\hspace*{0.02cm}2}} \sim
\begin{cases}
(t'^{\hspace*{0.02cm}2})^{\,-\frac{2U}{\pi\Gamma}  + O(U^2)} & V,\epsilon\hspace*{0.1cm} \ll~ t'\hspace*{0.07cm} \ll\Gamma \\
\hspace*{0.27cm}V^{\hspace*{0.1cm}\,-\frac{2U}{\pi\Gamma} + O(U^2)} & t',\epsilon\hspace*{0.1cm} \ll~ V \ll\Gamma \\
\hspace*{0.36cm}\epsilon^{\hspace*{0.15cm}\,-\frac{2U}{\pi\Gamma}  + O(U^2)} & V,t'\hspace*{0.02cm} \ll~ \epsilon\hspace*{0.15cm} \ll\Gamma~,
\end{cases}
\end{equation}
where higher-order corrections can again be extracted numerically. Such power-law behavior with respect to the voltage $V$ and hopping $t'$ was previously described using field-theoretical models\cite{doyon,dmrgnoneq} or perturbative renormalization group treatments,\cite{laszlo} and the corresponding exponents agree with the FRG result to leading order in $U$ (despite some difficulties in relating the parameters of a continuum model to our microscopic ones). Moreover, it is suggested in Ref.~\onlinecite{laszlo} that one can understand non-equilibrium properties from equilibrium only, which in the extreme limits mentioned above is supported by our observation that all bare system parameters (including the voltage) can be interpreted in terms of a simple infrared cutoff which automatically appear within the FRG framework (and do not have to be introduced by hand). In Sec.~\ref{sec:scaling.j}, we show that this altogether simple picture breaks down if the impurity energy is pinned to either one of the chemical potentials ($\epsilon\approx\pm V/2$).

\begin{figure}[t]
\includegraphics[height=5.1cm,clip]{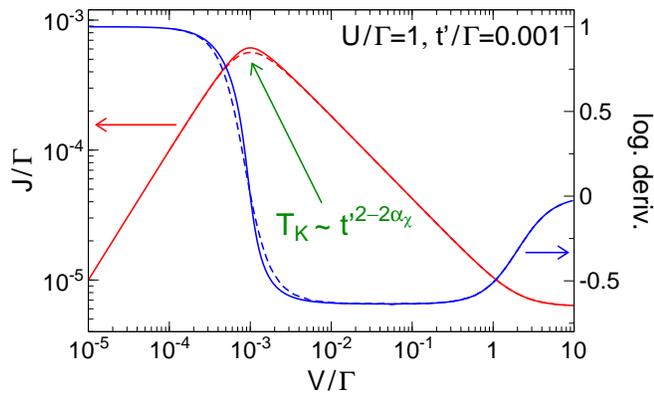}
\caption{(Color online) The current as a function of the voltage in the scaling limit $\Gamma\to\infty$ obtained numerically from the sharp (solid lines) and reservoir (dashed lines) FRG cutoff schemes, respectively. Beyond some cross-over scale $T_K$, $J$ decays as a power law of the voltage $V$ over several orders of magnitude. The latter manifests as a constant logarithmic derivative $d\ln(J/\Gamma)/d\ln(V/\Gamma)$.}
\label{fig:j}
\end{figure}

\subsection{Susceptibility}
\label{sec:scaling.chi}

In this Section, we investigate how the power laws described above manifest in equilibrium observables. Two quantities associated with linear-response transport are the resonance width of the conductance $G(\epsilon)$ and the charge susceptibility
\begin{equation}\label{eq:tk}
\chi = \frac{d\langle n\rangle}{d\epsilon}\Big|_{\epsilon=0} = -\frac{2}{\pi T_K}~,
\end{equation}
with $\langle n\rangle$ being the average occupation of the resonant level. The latter can conveniently be computed within the equilibrium FRG formalism by integrating over the Matsubara Green function. The inverse susceptibility can be used to define a unique scale $T_K$ which governs the low-energy linear-response physics.\cite{kondoscale} For small $U$, one can show that both the width of the conductance and $\chi$ are solely determined by the renormalized hopping $t'^{\la=0}$ computed in the previous Section and thus governed by a power law in the linear-response limit $V\ll T_K$:
\begin{equation}
\left(\frac{\chi^{-1}}{t'}\right)^2 \sim \left(\frac{t'}{\Gamma}\right)^{-2\alpha_\chi},~\alpha_\chi=\frac{2U}{\pi\Gamma} + O(U^2)~,
\end{equation}
where the higher-order corrections to the exponent are influenced by the flow of all self-energy components and can only be extracted numerically (by integration of Eqs.~(\ref{eq:floweqts}) - (\ref{eq:floweqes}) and subsequent power-law fitting; for the result see Fig.~\ref{fig:exp}). As mentioned above, one can in general not expect to obtain the right prefactor even of the second-order term since our truncated FRG scheme is correct only to leading $U$. However, going beyond first order allows both to specify the regime where the exponent is purely linear as well as for a quantitative comparison with other results.\cite{laszlo,doyon} In our case, the equilibrium numerical renormalization group can be straightforward employed to compute $\alpha_\chi$ with high accuracy, thus providing an additional benchmark for the functional RG in the scaling limit.\cite{nrg} One observes that the very simple (Hartree-Fock-like) FRG approximation scheme of Eqs.~(\ref{eq:floweqts}) - (\ref{eq:floweqes}) shows satisfying agreement with the NRG reference even for intermediate $U/\Gamma$ (see Fig.~\ref{fig:exp}).\cite{frgn1}

\subsection{Current}
\label{sec:scaling.j}

The most interesting transport property of the IRLM in non-equilibrium is the current. One can show that for small $U$ and large $\Gamma\gg t',V,\epsilon$ it is determined by the renormalized hoppings $\tau_{1,3}$, and an approximation to the latter can be derived from Eq.~(\ref{eq:flowts3}). This yields
\begin{equation}\label{eq:j}
J \approx \frac{4 \tau_{1}\tau_{3}}{\tau_{1}+\tau_{3}}\Bigg[
\arctan \left( \frac{V/2-\epsilon}{\tau_{1}+\tau_{3}}\right)+\arctan\left(\frac{V/2+\epsilon}{\tau_{1}+\tau_{3}} \right)\Bigg]~.
\end{equation}
In the following, we discuss this result as well as the numerically obtained current specifically for impurity energies in the middle between or close to either one of the chemical potentials of the bath ($\epsilon\ll V$ and $\epsilon\approx\pm V/2$, respectively).

\subsubsection{Zero impurity energy}

For large voltages $V\gg T_K, \epsilon$, Eq.~(\ref{eq:j}) describes a power law
\begin{equation}
\frac{J}{T_K} \sim \left(\frac{T_K}{V}\right)^{\alpha_J},~~ \alpha_J=\frac{2U}{\pi\Gamma} + O(U^2)~,
\end{equation}
with an exponent $\alpha_J$ that to leading order agrees with the results of Refs.~\onlinecite{dmrgnoneq}, \onlinecite{doyon}, and \onlinecite{laszlo}. Beyond the limit of small $U$, $\alpha_J$ as well as the current itself can only be computed numerically by integrating the flow equations (\ref{eq:flowsenoneq}) and (\ref{eq:flownoneqts1}) - (\ref{eq:flownoneqes}), respectively. Within both FRG non-equilibrium cutoff schemes, we observe that in agreement with our analytics $J$ generally features a linear increase crossing to a power-law decay (and thus a constant logarithmic derivative) at scale $V\approx T_K$\cite{tk} but eventually saturates as one approaches the bandwidth $\Gamma$ (see Fig.~\ref{fig:j}). The numerically-determined exponent $\alpha_J$ is depicted in Fig.~\ref{fig:exp}. It is purely linear up to sizable Coulomb interactions $U\approx\Gamma$, and our simple FRG approximation scheme does thus not contain higher-order corrections to the exponent in case of non-equilibrium.

As a passing comment, we note that in order to actually observe pure power-law behavior of the current, the voltage needs to be in a regime with $T_K\ll V\ll\Gamma$, and the bare hopping amplitude $t'$ typically has be chosen of the order of $t'^2/\Gamma\approx 10^{-6}$ (since $T_K\sim t'^2/\Gamma$ at small $U$). By successively increasing $t'$ we observe that the regime of voltages characterized by a power law shrinks until eventually for $t'=0.1\Gamma$ the logarithmic derivative only features a local minimum (with a value giving a rough estimate of the exponent, though) close to $V\approx\Gamma$. Within the DMRG framework of Ref.~\onlinecite{dmrgnoneq}, such large hoppings were used for a power-law fit of the current. Even though it is certainly numerically demanding, treating smaller $t'$ in non-equilibrium would be rewarded by putting the nice comparison of the DMRG results with a field-theoretical approach (which is a fundamental issue of Ref.~\onlinecite{dmrgnoneq}) on more solid grounds.

\begin{figure}[t]
\includegraphics[height=5.1cm,clip]{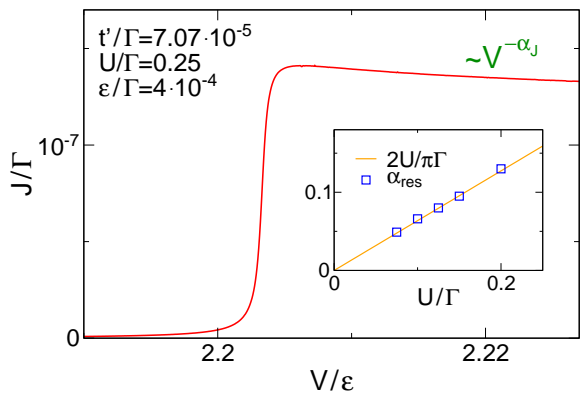}
\caption{(Color online) The same as Fig.~\ref{fig:j}, but for a finite impurity energy $\epsilon$. The data was obtained by numerically integrating the flow equations (\ref{eq:flownoneqts1}) - (\ref{eq:flownoneqes}). The current is suppressed for small voltages $V\ll\epsilon$ but crosses over to a power-law decay when $\epsilon$ is moved below the chemical potential.~\textit{Inset:} The quantity $\alpha_\tn{res}$ governing the behavior close to the aforementioned resonance condition $\epsilon\approx\pm V/2$ (see Eq.~\ref{eq:jres}). }
\label{fig:resonance}
\end{figure}

\subsubsection{On-resonance $\epsilon=\pm V/2$}

Quite intuitively, the current through the resonant level vanishes if the latter is energetically placed above (or below) of either one of the chemical potentials ($\epsilon\gg V$) while featuring the power-law decay described in the previous Section for small $\epsilon\ll V$ (see Fig.~\ref{fig:resonance}). If the impurity position is fixed to the so-called resonance condition $\epsilon=\pm V/2$ which determines the crossover between the aforementioned limits, the analytic expression (\ref{eq:j}) takes the form
\begin{equation}\label{eq:jres}
\frac{J}{T_K} \sim \frac{1}{1+\left(\frac{V}{T_K}\right)^{\alpha_{\tn{res}}}}~,~~\alpha_{\tn{res}}=\frac{2U}{\pi\Gamma} + O(U^2)~,
\end{equation}
for voltages $V\gg T_K$ where the arc-tangent can be replaced by its asymptotic value. However, even if $V$ is orders of magnitude larger than the equilibrium energy scale $T_K$, the current does not necessarily exhibit a power law since the exponent $\alpha_\tn{res}$ becomes small for $U\to0$. As before, we can verify this result numerically and extract the quantity $\alpha_\tn{res}$ beyond linear order by fitting to the form of Eq.~(\ref{eq:jres}).\cite{resonance} Within our (reservoir-cutoff) FRG approximation scheme, $\alpha_\tn{res}$ is equal to the off-resonance exponent $\alpha_J$ (see the inset to Fig.~\ref{fig:resonance}). Whereas those observations are altogether in complete agreement with recently-published real-time renormalization group results,\cite{irlmprl} they clearly contradict the intuition that the voltage can always be interpreted in terms of an infrared cutoff associated with universal power laws.\cite{doyon}

While for vanishing impurity energies both functional RG cutoff schemes describe the same non-equilibrium physics, the violation of causality within the sharp cutoff approach in presence of finite $\epsilon$ and voltages $V$ leads to severe artifacts (e.g., deviations from power-law behavior for $\epsilon\ll V$) already for fairly small Coulomb interactions. These violations originate from the diagonal self-energy components which do not flow at $\epsilon=0$. Thus, the reservoir cutoff scheme is clearly superior in investigating particle-hole asymmetric impurity positions for the IRLM beyond linear response.

\subsection{The left-right asymmetric IRLM}
\label{sec:scaling.asym}

So far, we have modeled the resonant level to be coupled symmetrically to both bath, and this scenario is generically employed in previous works.\cite{andrei,doyon,laszlo,dmrgnoneq} However, the FRG flow equations introduced in Sec.~\ref{sec:method} can be generalized straightforward for different Coulomb interactions $U_L\neq U_R$ and hopping amplitudes $t'_L\neq t'_R$ to the left and right side, respectively. Following the same arguing as in the symmetric case, one can analytically show that to leading order the current in the limit $\Gamma\gg V\gg t'_{L,R}, \epsilon$ takes the form
\begin{equation}
\frac{J}{T_K} \sim \frac{1}{\frac{1}{c}\left(\frac{T_K}{V}\right)^{-\frac{2U_L}{\pi\Gamma}} + c\left(\frac{T_K}{V}\right)^{-\frac{2U_R}{\pi\Gamma}} }~,
\end{equation}
where the linear-response low energy scale $T_K$ and the asymmetry parameter $c$ are given by $T_K\sim t'_Lt'_R$ and $c=t'_L/t'_R$ to zeroth order. For $U_L\neq U_R$, $J$ is thus not governed by a power law even if $V\gg T_K$ is large, and the voltage can again not be interpreted as an infrared cutoff. A more detailed discussion of the two-channel interacting resonant level model with asymmetric couplings can be found in Ref.~\onlinecite{irlmprl}.

\section{Conclusions}
\label{sec:conclusions}

In this work we have studied zero-temperature steady-state transport properties of the two-channel interacting resonant level model in presence of an arbitrary bias voltage $V$. Beyond linear response, the functional renormalization group in Keldysh frequency space can be used to compute the self-energy associated with the local Coulomb interaction between the isolated level and the two baths of delocalized states. We truncate the infinite hierarchy of functional flow equations to leading order, rendering the FRG an approximate method to calculate effective system parameters. Despite the simple (Hartree-Fock-like) nature of the resulting scheme, transport properties of the IRLM can be obtained to satisfying agreement with dynamical matrix renormalization group data both in and out of equilibrium. This allows for a thorough investigation of the special case where the bandwidth $\Gamma$ of the leads becomes large. For voltages much smaller than the linear-response energy scale $T_K$, this so-called scaling limit is characterized by universal power-laws. E.g., the charge susceptibility is governed by $\chi^{-1}\sim t'^{2-\alpha_\chi}\sim T_K$, with $t'$ being the local hopping to the leads. The exponent $\alpha_\chi=2U/\pi\Gamma+O(U^2)$ can be computed numerically and analytically and agrees with prior results to leading order. Far from equilibrium ($V\gg T_K$), the current decays as a power-law $J\sim V^{-\alpha_J}$ with the voltage if the impurity energy is small ($\epsilon\ll V$) while featuring more complex behavior if the latter is pinned close to either one of the chemical potentials of the bath ($\epsilon=\pm V/2$). Whereas in the former case the voltage can be interpreted in terms of a simple infrared cutoff (which automatically appears within the FRG framework), the same does not hold close to the resonance condition $\epsilon=\pm V/2$.

From the methodical point of view, we have established the functional renormalization group as a simple tool to compute effective (Hartree-Fock-like) parameters incorporating aspects of non-equilibrium physics of quantum impurity systems. The latter particularly holds for a recently-proposed way of implementing an infrared cutoff in Keldysh frequency space, which can be interpreted in terms of an additional reservoir whose coupling strength flows from infinity to zero and which does not suffer from symmetry violations specifically associated with non-equilibrium (such as causality). In general, however, the simple approximation obtained from truncating the infinite hierarchy of FRG flow equations is limited to treat small to intermediate values of the Coulomb interaction only. Extending the method to the strong-coupling limit is subject to future work.

\section*{Acknowledgments}
We thank N.~Andrei, B.~Doyon, P.~Schmitteckert, and A.~Zawadowski for fruitful discussions and particularly benefitted from joint work on this project together with S.~Andergassen, H.~Schoeller, and D.~Schuricht. The DMRG data was kindly provided by P.~Schmitteckert. We received support by the Deutsche Forschungsgemeinschaft via FOR 723 (CK, MP, and VM) and by the Alexander von Humboldt Stiftung (LB).

\begin{figure}[t]
\includegraphics[height=5.3cm,clip]{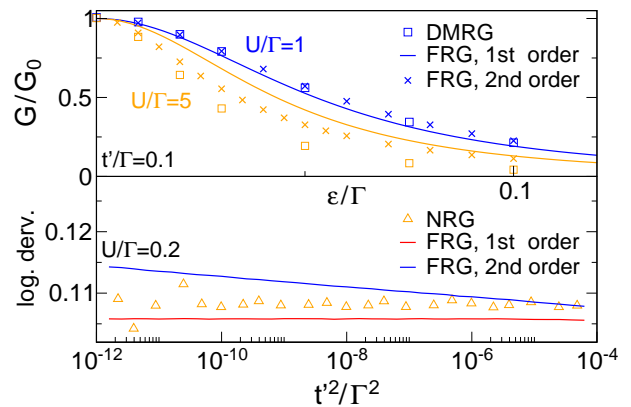}
\caption{(Color online) \textit{Upper panel:} The same as in Fig.~\ref{fig:dmrgg}, but additionally showing FRG results obtained from the second order truncation scheme outlined in the Appendix. \textit{Lower panel:} The logarithmic derivative $d\ln(\Gamma/\chi t'^{\,2})/d\ln(t'^{\,2}/\Gamma^2)$ obtained from the FRG flow equation (\ref{eq:floweqts}) as well as from a (frequency-independent) second-order generalization (see the main text for details) in comparison with numerical renormalization group data. Both for the NRG as well as for the first order FRG scheme, this quantity is constant over orders of magnitude as the charge susceptibility $\chi$ is governed by a power law of the bare hopping amplitude $t'$.}
\label{fig:secondorder}
\end{figure}

\appendix
\section*{Appendix: Second order functional RG}
\label{sec:appendix}

In this Section, we briefly discuss how the functional RG scheme employed to derive the results of Secs.~\ref{sec:dmrg} and \ref{sec:scalinglimit} can be extended by accounting for the flow of the two-particle vertex. Whereas we focus solely on the case of linear response, a detailed presentation of second-order FRG data in the context of the Anderson impurity model out of equilibrium can be found in Ref.~\onlinecite{severinsiam}.

Schematically, the flow equation of the two-particle vertex is given by\cite{dotsystems,katanin}
\begin{equation}\label{eq:flowga}
\partial_\la\gamma_2^\la \sim \tn{Tr } S^\la \gamma_2^\la G^\la \gamma_2^\la + \mathcal F\left(\gamma_3^\la\right)~,
\end{equation}
and after neglecting the contribution of the three-particle vertex $\gamma_3^\la$ (i.e., truncating the infinite hierarchy to second order) one obtains a finite set of differential equations which can in principle be solved numerically by introducing a discretization of Matsubara frequency space. For the single impurity Anderson model in equilibrium, this procedure (which is explained extensively in Ref.~\onlinecite{frequenzen}) leads to systematic improvements at small to intermediate Coulomb interactions and additionally allows for computing energy-dependent properties such as the local density of states or the finite-temperature conductance\cite{fermipaper} but cannot describe strong-coupling physics such as the appearance of the exponentially small Kondo energy scale.\cite{frequenzen}

For the problem at hand, one can pursue the same course of action and numerically solve the full (frequency-discretized) two-particle and self-energy flow equations (\ref{eq:flowga}) and (\ref{eq:flowse}), respectively. For the sake of numerical resources (and given the outcome) it is, however, desirable to devise some simplifications. Here, we only account for density-density (nearest and next-nearest neighbor) interaction terms but exemplary ensured that additionally considering the flow of `correlated hoppings' of the type $d_3^\dagger d_2^{\phantom{\dagger}} d_2^\dagger d_1^{\phantom{\dagger}}$ does not quantitatively alter our results for all cases of interest (particularly for the parameters of Fig.~\ref{fig:dmrgg} as well as for the equilibrium exponent $\alpha_\chi$ at $U\lesssim\Gamma$). Moreover, we approximate the frequency-dependence of the two-particle vertex by introducing three bosonic frequencies intrinsically connected to the three types of (particle-particle, particle-hole and hole-particle) diagrams appearing on the right-hand side of Eq.~(\ref{eq:flowga}) and discard all but these specific frequency dependencies (this is again outlined in more detail in Ref.~\onlinecite{frequenzen}).

As for the Anderson model, taking into account the flow of $\gamma_2^\la$ in the so-specified way leads to systematic improvements of physical quantities such as the linear-response conductance (see the upper panel of Fig.~\ref{fig:secondorder}). In addition, Ward identities (connected, e.g., to different equivalent ways of computing the average occupation number; see Ref.~\onlinecite{frequenzen}) are in general violated by the FRG (which is a non-conserving approximation) but fulfilled to much greater accuracy for intermediate Coulomb interactions within the second-order scheme. In contrast, the fundamental scaling-limit power-law behavior which is described by the differential equations (\ref{eq:floweqts}) - (\ref{eq:floweqes}) in complete agreement with NRG reference data is no longer captured if the flow of $\gamma_2^\la$ is accounted for (see the lower panel of Fig.~\ref{fig:secondorder}). This already manifests if the frequency-dependence of $\gamma_2^\la$ is completely discarded (and the second-order flow equation describes effective nearest- and next-nearest neighbor Coulomb interactions only). This is again analogous to the Anderson model where an exponential energy scale contained within the most simple (Hartree-Fock-like) approximation scheme\cite{dotsystems} is no longer captured by the (more elaborate) second-order approach.\cite{frequenzen,severinsiam} Thus, generalizing the FRG flow equations cannot be achieved by straightforward truncating at second instead of first order, and another strategy on how to tackle strong-coupling physics of quantum impurity systems using the functional renormalization group needs yet to be devised.



\begin{thebibliography}{*}
\bibitem{hartreefock}
In the context of quantum impurity systems, the FRG gives reasonable physical results despite its non-conserving nature, while the opposite often holds for Hartree-Fock calculations.
\bibitem{reservoirproblem}
The shortcoming of the `reservoir-cutoff' scheme introduced in Ref.~\onlinecite{severinsym} is that when implemented within the Matsubara formalism, second-order results are quantitatively inferior to those of the $\Theta$-function-approach. Namely, if one discards the flow of the three-particle vertex $\gamma_3^\la$ but fully accounts for the frequency-dependence of $\gamma_2^\la$, the former (the latter) gives reasonable results for the single impurity Anderson model in comparison with NRG data up $U/\Gamma=2$ ($U/\Gamma=6$), respectively (see Ref.~\onlinecite{severinsiam}). In order to tackle larger values of $U$ within the reservoir-cutoff approach, the second-order flow equations have to be further approximated (by neglecting, e.g., certain frequency-dependencies or self-energy feedbacks). For the problem at hand, we briefly present equilibrium second-order FRG results obtained from the sharp cutoff scheme only (see the Appendix).
\bibitem{gammaeq}
Similarly, one can show that in case of structureless leads implementing the reservoir cutoff scheme of Sec.~\ref{sec:method.gamma} directly within the Matsubara formalism gives the same analytic approximate (i.e., truncated to first order) self-energy flow equation as obtained using the sharp cutoff procedure.
\bibitem{logdiv}
One can show that perturbation theory to arbitrary order is completely regular even at zero frequency provided that $t'$ is finite.\cite{irlm4}
\bibitem{noneqcomment1}
In this case, we use the reservoir cutoff scheme only which allows for a simple interpretation of the self-energy as effective system parameters. Physical quantities (and the question whether they are governed by power laws in the scaling limit) will be investigated numerically using both non-equilibrium FRG frameworks (see Sec.~\ref{sec:scalinglimit}).
\bibitem{kondoscale}
In case of the single-channel IRLM, $T_K$ can be related to the Kondo temperature of the anisotropic Kondo model.\cite{irlm2,irlm4,qc}
\bibitem{nrg}
Standard NRG is a well-established numerical tool to compute low-energy equilibrium properties of quantum impurity systems. A detailed introduction to this method can be found in Ref.~\onlinecite{nrg3}.
\bibitem{frgn1}
For the single-channel IRLM, the same equilibrium exponent (which to leading order in $U$ agrees with $\alpha_\chi$) was computed from the functional and numerical renomalization group frameworks as well as by mapping the system to the anisotropic Kondo model by virtue of bosonization.\cite{qc} As for the two-channel case, the simple FRG approximation scheme employed in this paper agrees with the other approaches up to fairly large Coulomb interactions.
\bibitem{tk}
The equilibrium low energy scale $T_K$ can actually be defined as the voltage regime where the current crosses from a linear increase to a power-law decay (by considering, e.g., the maximum value of $J$). Except for prefactors, this definition is as expected equivalent to the one of Eq.~(\ref{eq:tk}), at least within the FRG approximation.
\bibitem{resonance}
In this case we neglect the renormalization of the level positions $\epsilon^\la$ and $\epsilon'^\la$. The resonance condition is then given by $\epsilon=\pm V/2$, and the tedious task of numerically determining the precise position $\epsilon_\tn{res}$ with an accuracy much smaller than the width $\Delta\epsilon_\tn{res}\sim T_K\ll\Gamma$ can be avoided. With the effects of the level renormalization being of higher order, this is justified if one only aims at verifying the analytic result of Eq.~(\ref{eq:jres}) and analyzing the trend how the quantity $\alpha_\tn{res}$ evolves beyond the linear regime.
\bibitem{katanin}
As for the Anderson model, we replace the single-scale propagator on the right-hand side of this flow equation by the cutoff derivative of the Green function.\cite{frequenzen} This is motivated by the fulfillment of Ward identities and changes our results only to the order of the truncation. However, both for the Anderson impurity and the interacting resonant level model discussed in this paper, this modified FRG scheme leads to quantitatively better results for intermediate Coulomb interactions (beyond second order).
\end{thebibliography}
\end{document}